# Defect induced negative magnetoresistance and surface state immunity in topological insulator BiSbTeSe$_2$

Karan Banerjee,[1] Jaesung Son,[1] Praveen Deorani,[1] Peng Ren,[2] Lan Wang,[2,3,*] and Hyunsoo Yang[1,†]

[1] Department of Electrical and Computer Engineering, NUSNNI, National University of Singapore, 117576 Singapore

[2] School of Physical and Mathematical Science, Nanyang Technological University, 637371 Singapore

[3] RMIT University, School of Applied Sciences, Department of Physics, VIC 3000, Australia

Absence of backscattering and occurrence of weak anti-localization are two characteristic features of topological insulators. We find that the introduction of defects results in the appearance of a negative contribution to magnetoresistance in the topological insulator BiSbTeSe$_2$, at temperatures below 50 K. Our analysis shows that the negative magnetoresistance originates from an increase in the density of defect states created by introduction of disorder, which leaves the surface states unaffected. We find a decrease in the magnitude of the negative magnetoresistance contribution with increasing temperature and a robustness of the topological surface states to external disorder.

Three-dimensional (3D) topological insulators (TIs) are a newly discovered state of matter with insulating bulk and conducting surfaces which are protected from backscattering by time reversal symmetry.[1-7] The band structure of a 3D TI consists of linearly dispersing surface states within the bulk band gap and is in the form of a spin-momentum locked Dirac cone which is a consequence of band inversion due to large spin-orbit coupling. The most commonly studied TI material to date is $Bi_2Se_3$ due to its relatively large bulk gap of 0.3 eV and favorable band structure, which allows experimentalists to access the topological surface states easily.[7] However, transport experiments using $Bi_2Se_3$ are plagued by large bulk conductivity due to the rapid formation of Se vacancies, which make it extremely challenging to access the physics of the surface states, since transport is overwhelmed by the bulk.[8-11] One of the possible ways to address this issue is to dope the material in order to decrease the concentration of Se vacancies and increase the insulating character of the bulk.[8, 12-16]

Magneto-transport measurements in TIs typically show weak anti-localization (WAL) behavior due to the large spin-orbit coupling and the linearly dispersing surface states.[17-20] The spin-momentum locking of the metallic surface states causes a suppression of backscattering which results in a π Berry phase acquired by the electrons executing time reversed paths. This results in destructive interference and observation of the WAL which is a quantum correction to the conductivity in an externally applied magnetic field. However, a few reports have previously demonstrated a signature of weak localization (WL) in TIs in the case of ultra-thin films in which the hybridization between the top and bottom surface wave-functions results in a gap opening in the surface states.[21-23] It has been also reported that magnetically doped TIs can exhibit WL.[24, 25] Given the large spin-orbit coupling and the absence of backscattering in the metallic surface states, the observation of WL in TIs is an interesting phenomenon, which is not fully understood.

Furthermore, the robustness of the surface states against external disorder has not been experimentally investigated.

In this work, we devise an experiment to introduce defects into the quaternary TI compound BiSbTeSe$_2$ (BSTS) by exposing the surface of the material to a beam of Ar$^+$ ions by using the ion milling process. We find that characteristic signatures of WL emerge in BSTS after ion milling. The signature of WL appears as a negative contribution to the magnetoresistance (MR) at temperatures lower than 50 K and at fields larger than 2 T. We also find that the topological surface states remain unaffected by the introduction of defects and maintain their immunity against disorder.

High quality BSTS single crystals used for the study was grown by the modified Bridgeman technique.[16, 26] In order to study the transport properties, photolithography was used to pattern thermally evaporated Cr (10 nm)/Au (90 nm) electrodes on the ~50−90 nm thick exfoliated BSTS nanoflakes. Transport measurements are performed in a 9 Tesla Quantum Design PPMS system. Argon ion milling was carried out on the samples at a power of 38 W and the distance between the source and the substrate was 30 cm.

Figure 1(a) shows a comparison of the resistance versus temperature plot between the different TI materials such as Bi$_2$Se$_3$ (inset) and BSTS. The Bi$_2$Se$_3$ flakes are exfoliated from polycrystalline Bi$_2$Se$_3$ powder (Alfa Aesar). Bi$_2$Se$_3$ shows the metallic behavior, whereas BSTS shows a non-metallic behavior. This indicates that the Fermi level lies inside the band gap in BSTS, whereas it lies inside the conduction band in Bi$_2$Se$_3$. It allows us to study the properties of the surface state with little interference from the bulk in the case of BSTS. In order to study the effect of ion milling, we first measure the transport properties of the BSTS control sample. The same sample is then subjected to ion milling and its transport characteristics are measured again to compare with the values before milling. Figure 1(b-d) shows the effect of ion milling on the transport characteristics of BSTS. We find that the corresponding gate voltage of the Dirac point has shifted from -40 to -80 V indicating that the sample has become more heavily n-doped after ion milling as shown in Fig. 1(b). The Boltzmann transport theory is used to fit the back gate voltage dependence of resistance as shown in Fig. 1(b).[11, 27, 28] From the Boltzmann fit, we obtain the charged impurity density ($n_{imp}$) of $7 \times 10^{12}$ and $1.33 \times 10^{13}$/cm$^2$ for the sample before and after milling, respectively, which indicates that the defect concentration has increased by almost two times after milling. The values around the Dirac point can be fitted more accurately by the effective medium theory, which is beyond the scope of the present study.

The carrier concentration increases by two times from $2.36 \times 10^{13}$ cm$^{-2}$ to $5.53 \times 10^{13}$ cm$^{-2}$ after milling as inferred from Hall measurements. We further find that electron mobility ($\mu_e$) decreases from 106 to 52 cm$^2$/V·s after milling. This is due to the increase in the density of defect states in BSTS after exposing it to ion milling. In order to detect the presence of the defect states in the band structure of BSTS, we have performed inelastic electron tunneling spectroscopy (IETS) measurements on our sample before and after milling at 4 K as shown in Fig. 1(c). IETS is the measurement of $d^2I/dV^2$ versus bias voltage (V) and is a very sensitive spectroscopic technique which has been used to study the detailed electronic structure.[29-31] It can hence provide information about the various inelastic electronic interactions and defect states present in case of a disordered semiconductor. It is seen that a defect peak appears at 8.6 meV in the IETS spectra, which becomes more distinct after milling. We also find that the intensity of the IETS signal increases significantly in the milled sample. An increase in the intensity of the IETS signal after milling indicates an increase in the density of defect states. Note that the defect states are present even in the control BSTS sample, but the density of the defect states increases after introduction of disorder, as inferred from Fig. 1(c).

Figure 1(d) shows the resistance versus temperature data for the sample before and after ion milling. It is interesting to note that the sample retains its non-metallic nature even after milling, indicating that the Fermi level lies inside the band gap after milling. We use an Arrhenius equation $R = R_0 \exp(E_a / k_B T)$ to fit the curve in Fig. 1(d), where $E_a$ corresponds to the activation energy.[18] We find that the activation energy for the control sample is 11.5 meV corresponding to a temperature of ~ 115 K. For the milled sample, activation energy is ~ 3.94 meV, corresponding to a temperature of ~ 40 K. When the temperature exceeds the activation energy, the electron transport is expected to be dominated by the bulk channels.

We now discuss the MR properties. The WAL phenomena in TIs are described by the Hikami-Larkin-Nagaoka (HLN) equation[32]

$$\Delta\sigma(B) = \frac{\alpha e^2}{\pi h} \left[ \psi\left(\frac{\hbar}{4eBL_\phi^2} + \frac{1}{2}\right) - \ln\left(\frac{\hbar}{4eBL_\phi^2}\right) \right]. \qquad (1)$$

In Eq. (1), $\psi$ is the digamma function, $B$ is the magnetic field, $\alpha$ is a prefactor, and $L_\phi$ is the phase coherence length. Each band that carries a $\pi$ Berry phase provides an $\alpha = -0.5$.[33, 34] In order to have a sharp WAL cusp, the phase coherence length must be larger than the electron mean free path.[35] The mean free path in the control BSTS sample is around 4.7 nm, as determined from a Fermi velocity $v_F = 5\times10^5$ m/s.[36] Fitting the MR data in Fig. 2(a) for the control sample with Eq. (1) at 2 K gives a value of $\alpha = -0.62$ and $L_\phi = 96$ nm which agrees well with previously reported values.[16, 19, 20, 37, 38]

The phase coherence length decreases as temperature increases due to increased thermal scattering, and when its value becomes less than the electron mean free path, the WAL cusp is lost.[37] Ordinarily, the MR contribution from the bulk remains negligible due to its unitary nature and the WAL signature in transport experiments comes predominantly from the surface states.[39] However, in the samples subjected to ion milling, a WL signature appears as a *negative contribution* to the MR at large fields as seen in Fig. 2(a-c). The appearance of negative MR after ion milling suggests an additional transport mechanism contributing to this unconventional behavior. Ion milling has been known to created defects like vacancies in semiconductors.[40-42] The process of ion milling is thus expected to increase the density of defect states in the BSTS band structure. The defect peak in the IETS spectra as shown in Fig. 1(c) confirms an increase in defect density in the BSTS sample after milling. Introduction of disorder results in the creation of localized states and electrons in these localized states behave differently from the conduction electrons.[43] The creation of additional defect states in the band structure competes with the linearly dispersed surface states in electron transport at low temperatures. At temperatures below 50 K, the electron transport occurs by hopping across the defect states.[43-46] When the electron transport due to hopping across the defect states becomes significant and starts competing with transport through surface states, it can result in changing the conventional MR behavior. The appearance of negative MR at high fields (> 2 T) thus can be attributed to electron hopping across the additional defect states, created by disorder.[22, 23, 47]

Note that the bulk maintains its non-metallic nature even after milling as seen in Fig. 1(d). This indicates that the Fermi level still lies inside the band gap. There have been previous reports[21-23] which suggest that the bulk channels in a topological insulator can give rise to WL in spite of the strong spin orbit coupling. However, this does not seem to be valid in our case since the bulk contribution for the milled sample becomes significant only above 40 K, as inferred from the Arrhenius fit in Fig. 1(d). Hence, it seems that the negative MR originates from the increase in the density of defect states created within the band gap due to ion milling, and not from the bulk channels. The negative MR appears as a parabolic superimposition upon the original WAL curve.

We use a summation of Eq. (1) and a parabolic field dependent term to fit the MR data from the milled sample,

$$\Delta\sigma(B) = \frac{\alpha e^2}{\pi h}\left[\psi\left(\frac{\hbar}{4eBL_\phi^2}+\frac{1}{2}\right)-\ln\left(\frac{\hbar}{4eBL_\phi^2}\right)\right]+\beta(H)^2 \qquad (2)$$

where $\beta$ is a co-efficient arising from the scattering across the defect states. We obtain good fits of the MR curves of the milled sample to Eq. (2), as shown in Fig. 2(d), implying that the negative MR indeed shows parabolic field dependence. Figure 3(a-b) shows the effect of ion milling on the fitting parameters $\alpha$ and the phase coherence lengths $L_\phi$ of the surface states obtained from the first term of Eq. (2) corresponding to the HLN equation. It is quite interesting to note that the $\alpha$ values and the phase coherence lengths of the BSTS sample are not affected much by ion milling, where the values for the control sample and milled sample are quite similar. The $\alpha$ value is close to -0.5 for both samples, as shown in Fig. 3(a). The phase coherence length $L_\phi$ is found to decay with temperature and shows a decaying power law trend scaling as $T^{-\gamma}$, as shown in Fig. 3(b). We obtain $\gamma$ = 0.58 and 0.56 for the sample before and after milling respectively, from the fits in Fig. 3(b). A value of $\gamma$ close to 0.5 indicates 2D electron-electron scattering for the sample before and after milling.[19, 20, 37] Hence, the electron transport is predominantly two-dimensional before and after milling. Furthermore, the electron dephasing time $\tau_\phi$ is found to be ~ 5 ns at 2 K, determined from the relation $L_\phi = \sqrt{D\tau_\phi}$, where $D = \mu_e k_B T/e = 1.82\times 10^{-6} m^2/s$, which is similar to the values extracted from recent reports on electron-electron interactions in topological insulators.[48-51] We observe that the behavior of the surface state of sample is relatively unaffected by the introduction of disorder, as inferred from the similar values of $\alpha$ and the behavior of $L_\phi$. This indicates that the surface states are remarkably robust against external damage induced by ion milling. The immunity against backscattering confers this protection to the surface states. The inset of Fig. 3(a) shows the evolution of the coefficient $\beta$ which shows a linear decrease in the magnitude, as the temperature changes from 2 to 50 K. The negative MR completely vanishes at 50 K when $\beta \sim 0$.

Negative MR has been previously reported and analyzed in disordered and doped semiconductors.[45, 52-54] It has been proposed that in disordered solids, there exists localized regions of spin accumulation.[43] On the application of an external field, these localized regions are aligned with the direction of the external field which results in a reduction of spin-scattering, leading to a decrease of the resistance with an increase in the field showing up as a negative contribution to MR. The negative MR shows a quadratic dependence with field in this case.[45] The quadratic negative MR originating from the defect states competes with the linear MR originating from the surface states. At large fields, the quadratic negative MR *overcomes* the linear positive MR and results in a downward bending of the MR curve as seen in Fig. 2(a-d). At low temperatures, the transport across the defect states is governed by the hopping probability across the states.[44, 46] The hopping probability decreases exponentially with an increase in temperature[44-46] resulting in a decrease in the magnitude of negative MR, consistent with our observations for the milled sample. We note that defect states are already present in the control sample as well, however the ion milling process significantly enhances the density of the defect states, resulting in an exponential increase in the hopping probability across the states. Therefore, the negative MR is visible only in the milled sample. Thus, we infer that the disorder introduced by ion milling appears as a negative contribution to MR by creating localized regions of spin accumulation which act as trapping sites for the electrons. At large fields, the alignment of the localized regions with the external field orientation results in a drop of resistance, which appears as a negative MR.

Figure 4(a) shows the effect of back gate bias at 2 K on the MR of the sample after milling, while the inset of Fig. 4(a) shows the gate bias effect on the MR before milling. We see that the MR is linear with a sharp WAL cusp across the Dirac point in the control sample before milling. In the milled sample in Fig. 4(a), which shows negative MR, the WL signature is found to become weaker as the Dirac point (~ -80 V) is approached. Application of gate bias at 30 K in the milled sample results in the disappearance of the negative MR signature and the MR percentage approaches that of the sample before milling as shown in Fig. 4(b). Thus, approaching the Dirac point restores the original WAL behavior of the surface states in the milled sample, demonstrating further the outstanding robustness of the surface states to disorder.

In summary, we observe the emergence of negative magnetoresistance in $BiSbTeSe_2$ subjected to ion milling by $Ar^+$ ions. The negative magnetoresistance appears at low temperatures and large magnetic fields, and disappears as the temperature exceeds 50 K. The origin of the negative MR is attributed to an increase in the density of defect states created by the introduction of disorder, leaving the surface states unaffected. The bulk channel contribution remains negligible even after milling and plays no major role in the negative MR behavior.

This work is supported by the Singapore Ministry of Education Academic Research Fund Tier 1 (R-263-000-A75-750) and Tier 2 (R-263-000-B10-112).

*WangLan@ntu.edu.sg

†eleyang@nus.edu.sg

FIG. 1. (a) Resistance (R) vs. temperature (T) curve of BSTS (the inset shows the metallic nature of $Bi_2Se_3$). (b) Gate voltage dependent resistance of BSTS with a Boltzmann fit before and after milling at 2 K. (c) Inelastic electron tunneling spectroscopy (IETS) of BSTS before and after milling at zero gate bias and 4 K. (d) Resistance vs. temperature curve of BSTS before and after milling with Arrhenius fits at zero gate bias.

FIG. 2. Out-of-plane magnetoresistance (MR) of BSTS before and after milling at 2 K with a fit (a), 10 K (b), and 30 K (c). (d) Magneto-conductivity of the milled BSTS sample at different temperatures and fits using Eq. (2).

FIG. 3. Fitting parameters $\alpha$ (a), $\beta$ (inset), and phase coherence lengths $L_\phi$ (b) as a function of temperature.

FIG. 4. (a) Effect of gate bias on magnetoresistance (MR) at 2 K after milling. The inset shows the effect of gate bias on MR at 2 K before milling. (b) Effect of gate bias on MR at 30 K after milling.

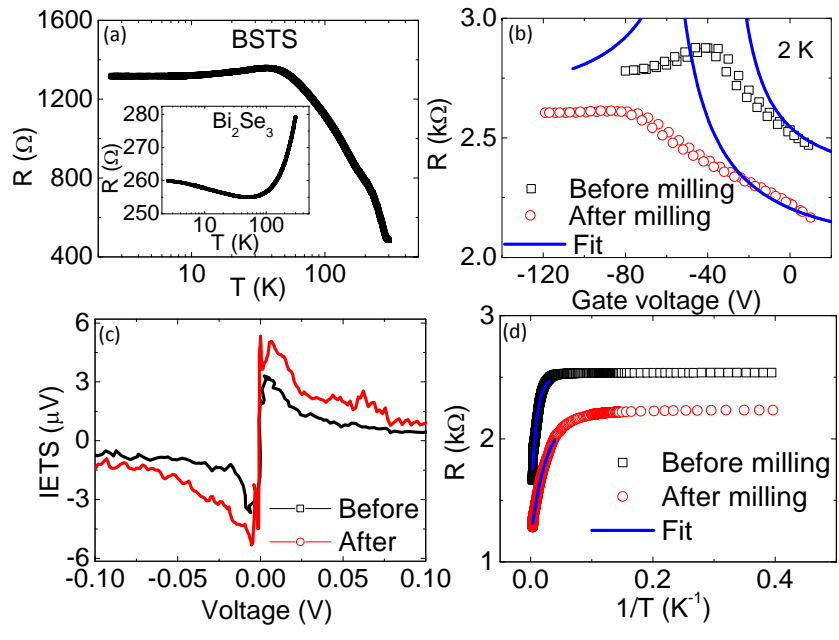

Figure 1

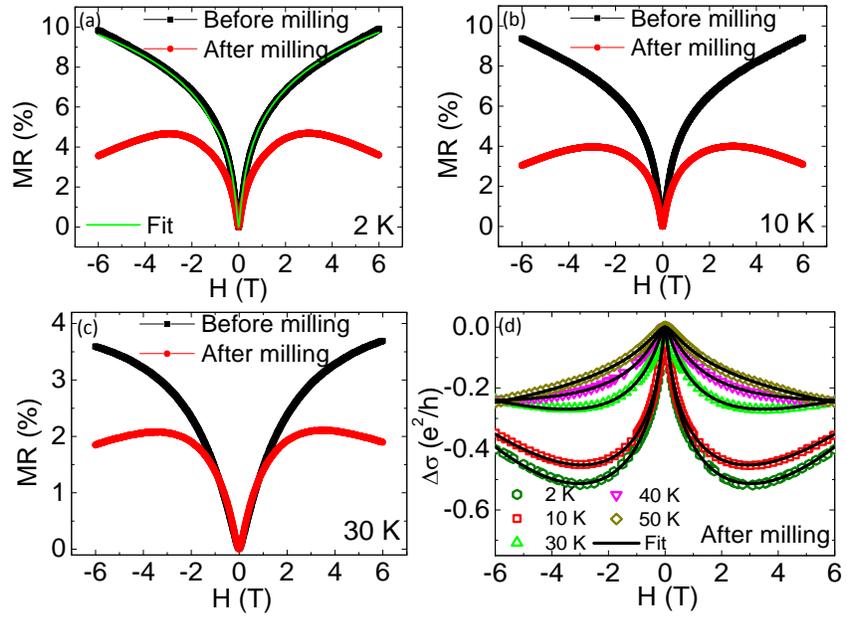

Figure 2

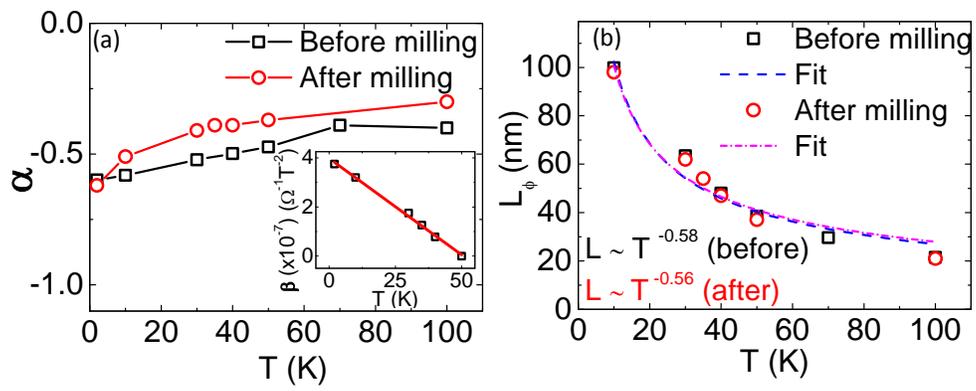

Figure 3

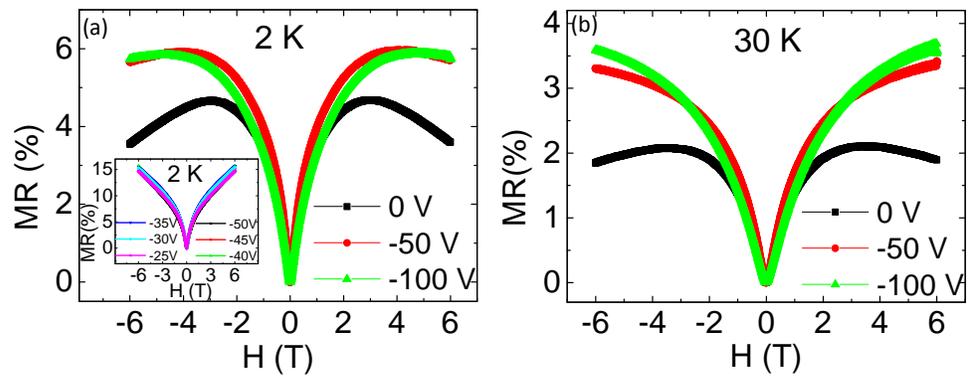

Figure 4